\documentclass[11pt]{article}
\usepackage{times}

\usepackage[colorlinks,linkcolor=blue,citecolor=blue,urlcolor=blue]{hyperref}
\usepackage{amsrefs}

\usepackage{fullpage,amsmath,amsfonts,amsthm}
\usepackage{url} 

\newtheorem{theorem}{\indent Theorem}

\newtheorem{corollary}[theorem]{\indent Corollary}

\renewcommand{\labelenumi}{(\arabic{enumi})}

\DeclareMathOperator{\tr}{tr}
\DeclareMathOperator{\poly}{poly}

\def\be{\begin{equation}}
\def\ee{\end{equation}}
\def\bea{\begin{eqnarray}}
\def\eea{\end{eqnarray}}
\def\ben{\begin{eqnarray*}}
\def\een{\end{eqnarray*}}

\def\benum{\begin{enumerate}}
\def\eenum{\end{enumerate}}
\def\bitem{\begin{itemize}}
\def\eitem{\end{itemize}}

\def\>{\rangle}
\def\<{\langle}

\newcommand{\bra}[1]{\mbox{$\left\langle #1 \right|$}}
\newcommand{\ket}[1]{\mbox{$\left| #1 \right\rangle$}}
\newcommand{\braket}[2]{\mbox{$\langle #1 | #2 \rangle$}}

\newcommand{\thmref}[1]{Theorem~\ref{thm:#1}}

\def\mn{\medskip\noindent}
\def\ot{\otimes}
\def\eps{{\epsilon}}
\def\cPU{\mathcal{PU}}
\def\cU{\mathcal{U}}
\def\bbC{\mathbb{C}}
\def\bbR{\mathbb{R}}
\def\bbZ{\mathbb{Z}}
\def\ra{\rightarrow}

\begin{document}
\title{Exact universality from any entangling gate without inverses}
\author{Aram W.\ Harrow\\
Dept. of Mathematics, U. of Bristol,
Bristol, BS8 1TW, U.K.\\
a.harrow@bris.ac.uk
}
\date{\today}
\maketitle

\begin{abstract}
This note proves that arbitrary local gates  together with any
entangling bipartite gate $V$ are universal.  Previously this was
known only when access to both $V$ and $V^\dag$ was given, or when
approximate universality was demanded.
\end{abstract}

A common situation in quantum computing is that we can apply only a
limited set $S\subset \cU_d$ of unitary gates to some $d$-dimensional
system.  The first question we want to ask in this situation is
whether gates from $S$ can (approximately) generate any gate in
$\cPU_d=\cU_d/\cU_1$ (the set of all $d\times d$ unitary matrices up to
an overall phase).  When this is possible, we say that $S$ is
(approximately) universal.  See \cites{DBE95,Lloyd95,Bryl02,FKL02} for
original work on this subject, or Sect 4.5 of \cite{NC00} or Chapter 8
of \cite{Kitaev:02a} for reviews.

Formally, $S$ is universal (for $\cPU_d$) if, for all $W\in \cPU_d$,
there exists $U_1,\ldots,U_k\in S$ such that
$$W = U_kU_{k-1}\cdots U_2U_1,$$
whereas $U$ is approximately universal (for $\cPU_d$) if, for all
$W\in \cPU_d$ and all $\eps>0$, 
there exists $U_1,\ldots,U_k\in S$ such that
\be d(W, U_kU_{k-1}\cdots U_2U_1) < \eps.\ee
Here $d(\cdot,\cdot)$ can be any metric, but for concreteness we will
take it 
to be the $\cPU_d$ analogue of operator distance:
$$d(U,V) := 1 - \inf_{\ket{\psi}\neq 0} \frac{\bra{\psi}U^\dag
  V\ket{\psi}}{\braket{\psi}{\psi}}.$$
Similar definitions could also be made for $\cU_d$, other groups, or
even semigroups.

A natural way to understand universality is in terms of the group
generated by $S$, which we denote $\<S\>$, and define to be smallest
subgroup of $\cPU_d$ that contains $S$.  An alternate and more
constructive definition is that $\<S\>$ consists of all products of a
finite number of elements of $S$ or their inverses.  When $S$ contains
its own inverses (i.e. $S=S^{-1}:=\{x:x^{-1}\in S\}$) then $\<S\>$
provides a concise way to understand universality: $S$ is universal
iff $\<S\>=\cPU_d$ and $S$ is approximately universal iff $\<S\>$ is
dense in $\cPU_d$.

But what if $S$ does not contain its own inverses?  The equivalence
between approximate universality and $\<S\>$ being dense in $\cPU_d$
still holds.  One direction remains trivial: if $S$ is approximately
universal then $\<S\>$ is dense in $\cPU_d$.  The easiest way to prove
the converse is with simultaneous Diophantine approximation,
which implies that for any $U\in \cPU_d$ and for any $\eps>0$, there
exists $n\geq 0$ such that $d(U^n,U^{-1})\leq \eps$.  The
proof is due to Dirichlet, and for completeness we include it
here\footnote{We prove the claim for $U\in\cU_d$, and the $\cPU_d$
  result will follow from the fact that ignoring a global phase can
  only decrease distance.
 Let the eigenvalues of $U$ be $(e^{2\pi
    i\alpha_1},\ldots, e^{2\pi i\alpha_d})$ for some $\alpha\in
  (\bbR/\bbZ)^d$.  Here $(\bbR/\bbZ)^d$ is the $d$-dimensional torus,
  which can be obtained by gluing together opposite faces of the
  hypercube $[0,1]^d$.  Note that under the $L_\infty$-norm, a ball of
  radius $\eps/2$ will have volume $\eps^d$.  Thus, if $n\geq
  1/\eps^d$ then the set $\{0,\alpha,2\alpha,\ldots,(n-1)\alpha\}$
  will have two distinct points, $n_1\alpha$ and $n_2\alpha$, with
  $\|n_1\alpha-n_2\alpha\|_\infty \leq \eps$.  If $n'=|n_2-n_1|$ then
  $0<n'<n$ and $\|n'\alpha\|_\infty\leq \eps$.  This implies that
  $\|U^{n'-1}-U^{-1}\|_\infty \leq |1-e^{i\eps}|=2\sin\eps/2 \leq
  \eps$.  }.  For any $W\in \cPU_d$ and $\eps>0$, the fact that $\<S\>$
is dense in $\cPU_d$ means that there exists an
$\frac{\eps}{2}$-approximation to $W$ of the form $U_1^{\pm 1}\ldots
U_k^{\pm 1}$, with each $U_i\in S$.  Now we replace each $U_i^{-1}$
term with $U_i^{n_i}$ for $n_i$ satisfying $\|U_i^{n_i}-U_i^{-1}\| \leq
\eps/2k$.  By the triangle inequality this yields an
$\eps$-approximation to $W$ out of a finite sequence of unitaries from
$S$.

The case of exact universality is more difficult, and is the subject
of the current note.  Again if $S$ is universal then $\<S\>=\cPU_d$,
and again we would like to argue that the converse holds.  Unfortunately
this statement is not known to be true, and there may well be
counter-examples 
along the lines of the Banach-Tarski paradox.  However in
the special case where $S$ contains a non-trivial one-parameter subgroup
then we can prove that universality with inverses implies universality
without 
inverses.  In fact we prove something a little stronger: not only can
any element of $\cPU_d$ be written as a finite product of elements from
$S$, but there is a uniform upper bound on the length of these
products.  If we define $S^L$ to be the set of products of $L$
elements from $S$, then we can prove
\begin{theorem}\label{thm:exact-univ}\ 
\renewcommand{\labelenumi}{(\alph{enumi})}
\benum
\item
  Suppose $S\subset \cPU_d$, $\<S\>=\cPU_d$ and there exists a 
  Hermitian matrix $H$ such that $H$ is not proportional to the
  identity and $e^{iHt}\in S$ for all $t\in\bbR$.
  Then $S$ is exactly universal for $\cPU_d$.  In fact there exists an
  integer $L$ such that $S^L=\cPU_d$.
\item 
  Suppose $S\subset \cU_d$, $\<S\>=\cU_d$ and there exists a 
  Hermitian matrix $H$ such that $H$ has nonzero trace, $H$ is not
  proportional to the 
  identity and $e^{iHt}\in S$ for all $t\in\bbR$.
  Then $S$ is exactly universal for $\cU_d$, and there exists $L$ such
  that $S^L=\cU_d$.
\eenum
\end{theorem}

The main interest of this theorem is in its application to the setting
of a bipartite quantum system where local unitaries are free and
nonlocal operations are restricted.  Say that $d=d_Ad_B$ and that
$S=\cU_{d_A}\times \cU_{d_B} \cup \{V\}$, where $\cU_{d_A}\times
\cU_{d_B}$ is embedded in $\cU_{d_Ad_B}$ according to $(U_A,U_B)\ra
U_A\ot U_B$ and $V$ is some arbitrary unitary in $\cU_{d_Ad_B}$.  In other
words, we can perform $V$ as well as arbitrary local unitaries,
meaning unitaries of the form $U_A\otimes U_B$.   Say that $V$ is {\em
  imprimitive} if there exists
$\ket{\varphi_A}\in\bbC^{d_A},\ket{\varphi_B}\in\bbC^{d_B}$ such that
$V(\ket{\varphi_A}\ot \ket{\varphi_B})$ is entangled.  Equivalently
$V$ is imprimitive if it cannot be written as $U_A\otimes U_B$ for any
$U_A\in\cU_{d_A},U_B\in\cU_{d_B}$, nor, if $d_A=d_B$, as
$\textsc{SWAP}\cdot(U_A\otimes U_B)$.  
Then \cite{Bryl02}
proved that $\<S\>=\cPU_d$ if and only $V$ is imprimitive.
 It was claimed in \cite{Bryl02} that
in fact $S$ was exactly universal when $V$ is imprimitive, but their
proof assumed that 
$V^\dag\in S$.  \thmref{exact-univ} then fills in the missing step in the proof
of \cite{Bryl02}, and together with the fact
that local unitaries contain at least one nontrivial one-parameter
subgroup and the results of \cite{Bryl02}, we obtain
\begin{corollary}
If $S=\cU_{d_A}\times \cU_{d_B} \cup \{V\}$ and $V$ is imprimitive
then $S$ is exactly universal for $\cU_{d_Ad_B}$.  In fact, there
exists an integer $L$ such that $S^L=\cU_{d_Ad_B}$.
\end{corollary}

This corollary is used in \cite{HS05} to prove that unitary gates have
the same communication capacities with or without the requirement that
clean protocols be used.  Exact universality there is used to show
that a protocol (possibly inefficient) exists for exact communication
using a fixed bipartite unitary gates supplemented by arbitrary local
operations. 
Now we turn to the proof of
\thmref{exact-univ}.
\begin{proof}
We start with an overview of the proof, and then discuss the details
of each step.  Let $G$ denote the group we are working with, which could
be either $\cPU_d$ or $\cU_d$, and let $m=d^2-1$ if $G=\cPU_d$ or $m=d^2$
if $G=\cU_d$.  Note that $G$ is an $m$-dimensional manifold\cite{GP74}.
\benum
\item We will define a smooth (i.e. infinitely differentiable) map $f$
  from $\bbR^m$ to $G$.  It will have the property that 
$df_0$ (its derivative at the point 0) is non-singular.
\item We will construct a map $\tilde{f}:\bbR^m\ra G$ such that
  $d\tilde{f}_0$ is non-singular and there exists an integer $\ell$ such
  that $\tilde{f}(x)\in S^\ell$ for all $x\in\bbR^m$.
\item We will construct an open neighborhood $N$ of the identity
  matrix $I\in G$ such that $N\subset
S^{\ell+\ell'}$ for some integer $\ell'$.
\item We will show that $G=N^n$ for some integer $n$, and thus that
  $G=S^{n(\ell+\ell')}$. 
\eenum

\mn{\em Step 1:} For some $U_1,\ldots,U_m\in G$ to be determined later,
we define
$$f(x) = U_1 e^{iHx_1} U_1^\dag 
U_2 e^{iHx_2} U_2^\dag \cdots U_m e^{iHx_m} U_m^\dag.$$
The partial derivatives at $x=0$ are given by
$$\frac{\partial f}{\partial x_j}(0) = iU_j H U_j^\dag.$$ We would
like to choose $U_1,\ldots,U_m$ so that the $U_j H U_j^\dag$ are
linearly independent.  Consider first the $G=\cPU_d$ case.  Then the
space of Hermitian traceless matrices (which we call
$\mathfrak{su}_d$) is a $d^2-1$-dimensional irrep of $G$, so the span
of $\{UHU^\dag:U\in G\}$ is equal to all of $\mathfrak{su}_d$.  Thus,
there exists a basis of $m=d^2-1$ matrices of the form $U_j
HU_j^\dag$.

When $G=\cU_d$, the tangent space is instead
the set of Hermitian matrices $\mathfrak{u}_d$, which decomposes into
irreps as $\mathfrak{u}_d = \mathfrak{su}_d \oplus \bbR I$.  Since $H$
is neither traceless nor proportional to $I$, it has nonzero overlap
with both irreps.
 Again we
would like to show that the span of $\{UHU^\dag:U\in G\}$ (which we
denote by $\mathfrak{h}$) is equal to
$\mathfrak{u}_d$.  First, we use the fact that $\cU_d$ acts
transitively on matrices of fixed spectrum.  Averaging over all $d!$
diagonal matrices isospectral to $H$ we obtain find that $(\tr
H)I/d$, which we
have assumed is nonzero, is in $\mathfrak{h}$.  Second, we replace $H$
with $H-(\tr H)I/d$ 
(which is in $\mathfrak{h}$ and $\mathfrak{su}_d$)
and use the result for $\cPU_d$ to show that the span of
$\mathfrak{su}_d\subset\mathfrak{h}$.  Thus $\mathfrak{h}$ equals all
of $\mathfrak{u}_d$.  Since $\mathfrak{h}$ was spanned by matrices of
the form $UHU^\dag$, this means we can  
choose a set of $d^2$ linearly independent matrices $U_1 HU_1^\dag,
\ldots, U_m H U_m^\dag$ to
form a basis for $\mathfrak{h}=\mathfrak{u}_d$.

In either case, $df_0$ has $m$ linearly independent columns of length
$m$, and thus is non-singular.  Denote the smallest singular
value of $df_0$ by $\eps$.

\mn{\em Step 2:}
Since $\<S\>=G$, $S$ is approximately universal and so we can
approximate $U_j$ and $U_j^\dag$ with products of elements of $S$, which we
call $\widetilde{U_j}$ and $\widetilde{U^\dag_j}$ respectively.
Demand that each approximation be accurate to within $\eps/4m$, so
that the total error is $\leq \eps/2$.  
We then define $\tilde{f}$ as follows:
$$\tilde{f}(x) := \widetilde{U_1} e^{iHx_1} \widetilde{U_1^\dag}
\widetilde{U_2} e^{iHx_2} \widetilde{U_2^\dag} \cdots 
\widetilde{U_m} e^{iHx_m} \widetilde{U_m^\dag}.$$
Note that $d(f(x),\tilde{f}(x))\leq \eps/2$ for all $x\in\bbR^m$.
Since $\eps$ is the smallest singular value of $df_0$, then the
smallest singular value of $d\tilde{f}_0$ must be $\geq \eps/2$, and
thus $d\tilde{f}_0$ is non-singular.

Additionally, each $e^{iHx_j}\in S$ and each $\widetilde{U_j}$ and
$\widetilde{U^\dag_j}$ is a product of a finite number of elements
from $S$, so there exists $\ell$ such that $\tilde{f}(x)\in S^{\ell}$ for all
$x\in\bbR^m$.

\mn{\em Step 3:} According to the inverse function theorem (see
e.g. \cite{GP74}), $\tilde{f}$ is a local diffeomorphism at 0.  This
means that there exists a neighborhood $X$ of 0 such that
$\tilde{f}(X)$ is a neighborhood of $\tilde{f}(0)$ and
$\tilde{f}:X\ra\tilde{f}(X)$ is a diffeomorphism (one-to-one, onto,
smooth and such that $\tilde{f}^{-1}$ is also smooth).  Let
$B_\delta(U):=\{V :d(U,V)< \delta\}$ denote the open ball of radius
$\delta$ around $U$.  Since $\tilde{f}(X)$ is a neighborhood of
$\tilde{f}(0)$, there exists $\delta>0$ such that
$B_{2\delta}(\tilde{f}(0))\subset \tilde{f}(X)$.  Now we again use the
approximate universality of $S$ to construct a $\delta$-approximation
to $\tilde{f}(0)^{-1}$, which we call $V$.  Then $V\cdot \tilde{f}(X)$
contains $B_\delta(I)=:N$.  Additionally, if $V\in S^{\ell'}$ then
$N\subset V\cdot \tilde{f}(X)\subset S^{\ell + \ell'}$.

\mn{\em Step 4:} If $n>\pi/2\sin^{-1}(\delta/2)$ then
$B_\delta(I)^n=G$. This is because $G=\{e^{iH}:\|H\|_\infty \leq
  \pi\}$ (optionally modulo overall phase) and $B_\delta(I) =
  \{e^{iH}:\|H\|_\infty \leq 2\sin^{-1}(\delta/2)\}$.
 Thus $G=S^{n(\ell+\ell')}$.
\end{proof}

We conclude with some open questions. First, it would be nice to know
the exact conditions on $S$ for which $\<S\>=G$ implies exact
universality.  A perhaps more important question is that of
efficiency.  If $S$ is approximately universal and contains its own
inverses, then the Solovay-Kitaev theorem\cites{Kitaev:02a,Dawson05}
states that any gate can 
approximated to an accuracy $\eps$ by $S^\ell$ for
$\ell=\poly\log(1/\eps)$.  But if $S$ does not contain its own
inverses, the best bound known on $\ell$ is the trivial
$\poly(1/\eps)$ bound from Dirichlet's theorem.

{\em Acknowledgments:} My funding is from the U.S. Army Research Office
under grant W9111NF-05-1-0294, the European Commission under Marie
Curie grants ASTQIT (FP6-022194) and QAP (IST-2005-15848), and the
U.K. Engineering and Physical Science Research Council through ``QIP
IRC.''


\bibliographystyle{alpha}

\end{document}